\documentstyle[preprint,prl,aps]{revtex}

\begin{document}
\def\sn2{$\sin^22\theta$}
\def\dm2{$\Delta m^2$}
\def\ch2{$\chi^2$}
\def\moko{\stackrel{<}{\sim}}
\draft
\begin{titlepage}
\preprint{\vbox{\hbox{IASSNS-AST 94/54}
\hbox{UDHEP-10-94}
\hbox{hep-ph 9410353} \hbox{October 1994} }}
\title{ \large \bf Solar Neutrinos and the Principle of Equivalence} 
\author{\bf J.N. Bahcall$^{(a)}$, P.I. Krastev$^{(a)}$
\footnote{Permanent address:
Institute of Nuclear Research and Nuclear Energy, Bulgarian Academy
of Sciences, BG--1784 Sofia, Bulgaria.} 
and C.N. Leung$^{(b)}$}
\address{(a) School of Natural Sciences, Institute for Advanced Study\\
Princeton, NJ 08540\\}
\address{(b) Department of Physics and Astronomy, University of Delaware\\
Newark, DE 19716 \\}
\maketitle
\begin{abstract}

We study the proposed solution of the solar neutrino problem which
requires a flavor nondiagonal coupling of neutrinos to gravity.  We
adopt a phenomenological point of view and investigate the
consequences of the hypothesis that the neutrino weak interaction
eigenstates are linear combinations of the gravitational eigenstates
which have slightly different couplings to gravity, $f_1G$ and $f_2G$,
$|f_1-f_2| << 1$, corresponding to a difference in red-shift between
electron and muon neutrinos, $\Delta z/(1+z) \sim |f_1 - f_2|$.  We
perform a $\chi^2$ analysis of the latest available solar neutrino
data and obtain the allowed regions in the space of the relevant
parameters.  The existing data rule out most of the parameter space
which can be probed in solar neutrino experiments, allowing only $|f_1
- f_2| \sim 3 \times 10^{-14}$ for small values of the mixing angle
($2 \times 10^{-3} \le \sin^2(2\theta_G) \le 10^{-2}$) and $10^{-16}
\stackrel{<}{\sim} |f_1 - f_2| \stackrel{<}{\sim}10^{-15}$ for large
mixing ($0.6 \le \sin^2(2\theta_G) \le 0.9$).  Measurements of the
$^8{\rm B}$-neutrino energy spectrum in the SNO and Super-Kamiokande
experiments will provide stronger constraints independent of all
considerations related to solar models.  We show that the
recoil-electron spectrum measured by the Kamiokande-II collaboration
can be used to exclude part of the allowed regions obtained above.  We
analyze the prospects of using future spectral measurements of solar
neutrinos to distinguish the oscillation mechanism due to the
violation of the equivalence principle from the conventional
mechanisms which require neutrinos to have nondegenerate masses.  We
find that, for small mixing angles, these mechanisms lead to very
different spectral predictions which will be distinguishable in the
upcoming SNO and Super-Kamiokande experiments.

\end{abstract}
\end{titlepage}

\newpage
\section{Introduction}

Results from four solar neutrino experiments\cite{CL}-\cite{GALLEX}
utilizing different detection techniques consistently show a
discrepancy between the measured $\nu_e$ flux from the sun and the
$\nu_e$ flux predicted by various solar models\cite{BP}-\cite{TC}.
Moreover, a comparison of two of the experiments suggests that,
essentially independent of solar models, some new physical process may
be changing the results of the neutrino energy spectrum\cite{BB}.  The
origin of this solar neutrino deficit is not yet known. A possible
solution is neutrino flavor oscillations.  The mechanism for neutrino
oscillation originally proposed by Pontecorvo\cite{Pontecorvo} assumes
that neutrinos have nondegenerate masses and that the neutrino mass
eigenstates are distinct from their weak interaction eigenstates.  We
shall refer to this as the mass mechanism.  The most often discussed
version of this type of solutions is the Mikheyev-Smirnov-Wolfenstein
(MSW) effect\cite{MSW}, in which the solar electron neutrinos can be
converted almost completely into muon or tau neutrinos due to the
presence of matter in the Sun.

An alternative mechanism of neutrino oscillation which does not
require neutrinos to have a nonzero mass was first proposed by
Gasperini
\cite{Gasperini}, and independently by Halprin and Leung\cite{HL}, 
as a means to test the 
equivalence principle (EP).  In this mechanism, neutrino oscillations
occur as a consequence of an assumed flavor nondiagonal coupling of
neutrinos to gravity which violates the EP.  We shall refer to this as
the VEP mechanism.

The VEP mechanism has been studied further in a number of
papers\cite{PHL}--\cite{MN}.  One study\cite{Butler} claims that this
solution is already excluded by the data from the four solar neutrino
experiments.  On the other hand, a $\chi^2$ analysis of the data
performed in another study\cite{PHL} reveals regions of the parameter
space which are allowed by the results of all four solar neutrino
experiments.  One of our aims in the present paper is to address this
controversy.  We repeat the analysis using the most current solar
neutrino data and find that, although strongly restricted by the data,
this VEP mechanism is a phenomenologically allowed solution of the
solar neutrino problem and deserves further study.

Having established the VEP mechanism as a possible solution to the
solar neutrino problem, it is necessary to find ways to determine
whether the mass or the VEP mechanism is responsible for the observed
solar neutrino deficit.  We show that, because of their differing
energy dependence, the two mechanisms yield, in the case of small
mixing angles, very different and distinguishable predictions for
spectral measurements in future solar neutrino experiments.

\section{Formalism and Solutions}

We begin by recalling the important features of the two oscillation
mechanisms.  For simplicity, we only consider mixing of two neutrino
flavors, e.g., $\nu_e$ and $\nu_\mu$.  In the mass mechanism, the
neutrino weak interaction (flavor) eigenstates, $\nu^{(W)} = (\nu_e,
\nu_\mu)$, are assumed to be linear superpositions of the mass
eigenstates, $\nu^{(M)} = (\nu^{(M)}_1, \nu^{(M)}_2)$, with a mixing
angle $\theta$.  The evolution equations for relativistic flavor
neutrinos propagating in vacuum are given by
\begin{equation}
i \frac{d}{dr}\left( \begin{array}{c}
\nu_e \\
\nu_\mu \end{array}
\right) = \frac{\Delta m^2}{2 E} \left[ \begin{array}{cc}
0  &  \frac{1}{2} \sin 2\theta \\
\frac{1}{2} \sin 2\theta & \cos 2\theta \end{array}
\right] \left( \begin{array}{c}
\nu_e \\
\nu_\mu \end{array}
\right),
\label{evol}
\end{equation}
where $E$ is the neutrino energy, $\Delta m^2 \equiv m_{2}^2 -
m_{1}^2$, and $m_{1,2}$ are the mass eigenvalues.  The survival
probability for a $\nu_e$ after traveling a distance $L$ is
\begin{equation}
P(\nu_e \rightarrow \nu_e) = 1 - \sin^2 2\theta \sin^2 \frac
{\pi L}{\lambda_M}, 
\label{surv}
\end{equation}
where $\lambda_M$ is the oscillation length defined as 
\begin{equation}
\lambda_M = \frac{4 \pi E}{\Delta m^2}.
\label{lm}
\end{equation}

In the VEP mechanism neutrinos are assumed to be massless and there is
no mass--dependent mixing.\footnote{We do not consider explicitly the
case of massive neutrinos in the VEP mechanism because this just
complicates the analysis by introducing more parameters, but does not
change the fundamental limits that are derived.  See Ref.\cite{MN} for
a recent discussion of this case.}.  Instead the weak interaction
eigenstates are assumed to be linear superpositions of the
gravitational interaction eigenstates, $\nu^{(G)} = (\nu^{(G)}_1,
\nu^{(G)}_2)$, with a mixing angle $\theta_G$.  It is further assumed 
that $\nu^{(G)}_1$ and $\nu^{(G)}_2$ interact with gravity with
slightly different couplings, thus violating the EP and leading to
neutrino oscillations when a neutrino propagates in a gravitational
field.  The neutrino evolution equations and the survival probability
for a $\nu_e$ after traversing a distance $L$ in the weak
gravitational field of a static, spherical symmetric source are given
(in the harmonic gauge) by Eqs.~(\ref{evol}) and (\ref{surv}), with the
substitutions\cite{Gasperini}\cite{HL},
\begin{equation}
\theta \rightarrow \theta_G~~~~{\rm and}~~~~\frac{\Delta m^2}{2E} 
\rightarrow 2E |\phi(r)| \Delta f ,
\label{sub}
\end{equation}
where $\phi(r)$ is the Newtonian gravitational potential and 
$\Delta f \equiv f_2 - f_1$ is a measure of the degree of 
EP violation.  Here $f_{1,2}$ can be identified as parameters 
in the parametrized post-Newtonian formalism\cite{Korea}\cite{HLP}, 
and $f_1 = f_2$ if the EP is obeyed.  In general relativity, 
$f_1 = f_2 = 1$. 

There has been much discussion in the literature \cite{pot}, mostly in
the context of testing the EP in the $K^0-\overline{K^0}$ system, as
to whether the gravitational potential in Eq.~(\ref{sub}) should include 
the potential of matter other than the Sun.  If distant matter is
included, then one obtains the counterintuitive result that the
biggest effect is from material far away from the solar system.  We
adopt here the phenomenological point of view that the potential
$\phi(r)$ is the difference between the Newtonian potential of the
solar material at $r$ and at $r = \infty$.

It is conceivable that the description of the violation of the
equivalence principle outlined above will be modified if the violation
is derived from a more rigorous theory, e.g., from string theory
\cite{polya}. The dependence on the gravitational potential might be
replaced in such a thory by a dependence on the
gradient of the potential, which would eliminate the above-mentioned
ambiguity concerning the relevant gravitational potential.  For example, 
the term in Eq.~(\ref{sub}) may be replaced by 
\begin{equation}
2E |\phi(r)| \Delta f \rightarrow E {\rm R}_f |\nabla\phi| 
\label{eq:grad}
\end{equation}
\noindent
where ${\rm R}_f$ is a dimensional parameter that describes the violation 
of the equivalence principle.  The results obtained below for the VEP 
mechanism as defined in Eq.~(\ref{sub}) can be easily translated to the case 
represented by Eq.~(\ref{eq:grad}).  Since 
$|\nabla\phi| \sim R^{-1}_{\odot} |\phi|$, 
where $R_{\odot}$ is the solar radius, we can interpret the
numerical results for $\Delta f$ in Eq.~(\ref{sub}) as constraints on a
mass scale for flavor violation, 
\begin{equation}
{\rm M}_f \simeq (R_{\odot} \Delta f)^{-1}
\label{eq:scale}
\end{equation}

It follows from the substitution (\ref{sub}) and Eq.~(\ref{lm}) 
that the oscillation length in the VEP mechanism is, for a 
constant gravitational potential,
\begin{equation}
\lambda_G = \frac{\pi}{E |\phi| \Delta f}
\label{lg}
\end{equation}
Notice that $\lambda_G$ is inversely proportional to the neutrino 
energy, whereas $\lambda_M$ increases with $E$.  
It is this different energy dependence that leads to the observable 
distinction between the VEP and mass mechanisms.

Another consequence of this differing energy dependence is that, in
contrast to the mass mechanism, the VEP mechanism cannot account for
the observed solar neutrino deficit as the result of long-wavelength
vacuum oscillations\cite{PHL}.  On the other hand, the solar neutrino
data can be encompassed through the MSW mechanism\cite{MSW} of
resonance enhanced transitions in the sun\cite{PHL}.  For neutrinos
propagating in matter, the evolution equations, Eq.~(\ref{evol}), for
the mass (VEP) mechanism are modified such that
\begin{equation}
\cos 2\theta_{(G)} \rightarrow \cos 2 \theta_{(G)} - 
\frac{\sqrt{2} G_F N_e(r)}{2 \pi/\lambda_{M(G)}}  
\label{matter}
\end{equation}
where $G_F$ is the Fermi constant and $N_e(r)$ is the 
electron number density inside the Sun.  In the VEP case, the 
resonance occurs when 
\begin{equation}
E = \frac{\sqrt{2}G_F N_e(r)}{2 |\phi (r)| \Delta f \cos 2 \theta_G}
\label{eq:res}
\end{equation}

An important ingredient in the analysis of resonant transitions in the
Sun is the adiabaticity condition.  For the VEP mechanism, it reads
\begin{equation}
\kappa = 
\frac{\sqrt{2}G_F\left(N_e\right)_{res}\tan^22\theta_G}{\left|
\left(\frac{1}{N_e}\frac{dN_e}{dr}\right) -
\left(\frac{1}{\phi}\frac{d\phi}{dr}\right)\right|_{res}} \gg 1.
\label{eq:adia}
\end{equation}
{}For the mass mechanism, simply drop the $\phi^{-1}d\phi/dr$ term in
the denominator and replace $\theta_G$ by $\theta$.  The dependence on
the energy is implicit as the energy determines the resonant density
via Eq.~(\ref{eq:res}).  Note that the adiabaticity condition in the
VEP case is violated for low energy neutrinos whereas it is violated
in the mass mechanism for high energy neutrinos.

It can be shown from Eqs.~(\ref{evol}) and (\ref{matter}) that the
probability amplitude ($A_e(r)$) of finding an electron neutrino at a
distance $r$ from where it was produced satisfies the equation,
\begin{equation}
A''_e + A'_e (ia - b'/b) + b^2 A_e = 0,
\label{eq:evoleq2}
\end{equation}
where the prime symbols denote derivatives with respect to $r$.  
The parameters $a$ and $b$ are defined as
\begin{equation}
a = -\sqrt{2}G_F N_e(r) - 2 E \phi (r) \Delta f \cos 2\theta_G,
\end{equation}
\begin{equation}
b = - E \phi (r) \Delta f \sin 2\theta_G.
\end{equation}
Eq.~(\ref{eq:evoleq2}) is exactly solvable in the case of a Newtonian
gravitational potential and zero electron density (e.g., outside the
Sun).  The general solution, expressed in terms of the dimensionless
variable, $x = r/R_{\odot}$, has the form
\begin{equation}
A_e(x) = C_1 x^{iS\cos^2\theta_G} 
       + C_2 e^{i\omega}x^{-iS\sin^2\theta_G},
\label{eq:gsol}
\end{equation}
where $C_1, C_2$ and $\omega$ are real constants that have to be
determined by the initial conditions.  For a neutrino moving in the
gravitational potential of the Sun, $S = 2 E \Delta f G_N M_{\odot}$,
where $G_N$ is Newton's gravitational constant and $M_{\odot}$ is the
solar mass.  The probability to find a $\nu_e$ at a distance $x$, if
at $x$ = 1 a $\nu_e$ has been produced, is thus
\begin{equation}
P(x) = 1 - \sin^2(2\theta_G) \sin^2(\frac{S}{2} \ln(x)),~~~~~~x \geq
1.
\label{eq:prob}
\end{equation}
By analogy with neutrino oscillations in vacuum one can introduce
the oscillation length, ${\rm L}_G = 2\pi x R_{\odot}/S\ln(x)$, which
turns out to be distance dependent. From (\ref{eq:gsol}) one can
derive the average probability to obtain an electron neutrino at
infinity given an arbitrary neutrino state at $r = R_{\odot}$:
\begin{equation}
\bar{P} = 
\cos^22\theta_G P(1) - \sin2\theta_G\cos2\theta_G R(1) 
+ \frac{1}{2}\sin^22\theta_G.
\end{equation}
Here $R(1) = Re[A_e(1)A^*_{\mu}(1)]$.  This expression is necessary
for the computation of the mean survival probabilities, especially in
the case of large mixing angles, $\sin^22\theta_G\ge 0.1$, where one
has to average over large-amplitude oscillations in vacuum between the
Sun and the Earth.  It is important also for the analysis in the
long-wavelength regime where the oscillation length becomes
comparable to the Earth-Sun distance.  We have used these results to
verify the finding in \cite{PHL} that the possibility of a
long-wavelength (or ``just-so'') VEP solution to the solar neutrino
problem is ruled out by the present data.

The survival probability as a function of the product $E \Delta f$ is
shown in Fig.~1 for a small as well as a large mixing angle.  The
curves in Fig.~1 have been obtained after averaging over the neutrino
production regions \cite{BP} of the different components of the solar
neutrino flux. The survival probabilities have been computed using the
analytical result,
\begin{equation}
P = \frac{1}{2} + ( \frac{1}{2} - P_{LZ}
)\cos2\theta_G^o\cos2\theta_G,
\label{eq:psur}
\end{equation}
where 
$P_{LZ} = (e^{-\beta} - e^{-\alpha})/(1 - e^{-\alpha})$, 
with $\alpha = 2\pi\kappa\cos2\theta_G/\sin^22\theta_G$ 
and $\beta = \frac{\pi}{2}\kappa ( 1 - \tan^2\theta_G )$.  
Here $\theta_G^o$ is the mixing angle at the production point of the
neutrino.  These expressions have been obtained by analogy with the
ones in \cite{serg} for MSW-transitions in the mass mechanism.  It is
exact in the case of a density which varies exponentially with the
distance, $r$, assuming that the variation of the gravitational
potential with distance is much slower than the variation of the
density with distance and can be neglected altogether. For small
mixing angles this is an excellent approximation as the scale height
of the gravitational potential is much larger than the density scale
height.  For example, the gravitational potential changes by a factor
less than ten from the center of the Sun to the surface whereas the
density changes by several orders of magnitude. We use the density
distribution inside the Sun as given in
\cite{BP} both for the electron number density and to compute the
gravitational potential inside the Sun. For each value of $E \Delta f$
the gravitational potential has been put equal to the value it assumes
at the point where the resonance takes place.  We have verified by
numerical integration of the evolution equations that the results
obtained with the analytical formula (\ref{eq:psur}) are accurate to a
few percent. In contrast with the mass mechanism, the adiabatic edge
of the suppression pit in the case of VEP is at higher energies and,
for small mixing angles, is bounded by the maximal density in the sun
($\approx 100 N_A$ cm$^{-3}$) to be at $E \Delta f \approx 10^{-12}$
MeV.

The resonance and adiabaticity conditions together determine the range
of parameters which can be probed by solar neutrino experiments. This
is depicted in Fig.~2 as the region bounded by the dotted lines. The
horizontal line corresponds to a resonant density equal to the density
in the center of the Sun for neutrinos with energy 0.2~MeV.  For
$\Delta f > 2\times10^{-12}$ there will be no resonance crossing in
the Sun for neutrinos with energies higher than 0.2 MeV. Oscillations
with an amplitude $\sim \sin^22\theta_G$ will still take place,
however, the oscillation length at the Earth for neutrinos of energies
0.1 MeV and higher will be smaller than $10^{-4} {\rm
R_{\odot}}$. Therefore the averaging over the distance between the
source and detector will result in energy independent suppression of
all solar neutrino fluxes, which does not give an acceptable fit to
the data.  The diagonal line in Fig.~2 corresponds to $\kappa = 1$ for
neutrinos of energy 10~MeV.  For values of $\sin^2 2\theta_G$ and
$\Delta f$ in the region below this line the transitions are strongly
nonadiabatic and cannot account for the solar neutrino problem.

\section{Comparison with data}

In section III-A we show which regions of the parameter space for the
VEP mechanism are ruled out by the measured counting rates in the four
operating solar neutrino experiments. In section III-B we use the
implied spectral distribution of the $^8$B-neutrino spectrum and the
existing Kamiokande measurements to further reduce the allowed
parameter space. We also show in Section III-B, and especially Figures
3 and 4, how future spectral measurements with SNO and
Super-Kamiokande can be used to distinguish between different
mechanisms for solving the solar neutrino problem.

\subsection{Rates}

We use the following recent experimental results: $ Q_{Cl} = (2.55 \pm
0.25) SNU$ \cite{CL}, $\Phi_{K}(^8B) = (2.89 \pm 0.41) \times 10^6~
{\rm cm}^{-2}$s$^{-1}$ \cite{KII}, $Q_{Ga} = (73 \pm 19.3) SNU$
\cite{SAGE}, and $Q_{Ga} = (79 \pm 11.7) SNU$ \cite{GALLEX}.

In our $\chi^2$ analysis of the latest solar neutrino data in terms of
the ``MSW-enhanced" VEP mechanism we have adopted a procedure that
takes into account the theoretical uncertainties in the standard solar
model as described in \cite{hatal}. The analysis yields two allowed
regions at 95~\% C.L., a ``small mixing region'' for: 
$2 \times 10^{-3} \le \sin^2(2\theta_G) \le 10^{-2}$  and 
$2.7 \times 10^{-14} \le \Delta f \le 3.3 \times 10^{-14}$; 
and a ``large mixing region'' for: 
$0.6 \le \sin^2(2\theta_G) \le 0.9$ and 
$1.0 \times 10^{-16} \le \Delta f \le 1.5 \times 10^{-15}$.  
These are shown in Fig.~2 by the unhatched regions within the solid curves.  
The quality of the fit is better for the small mixing solution where
$\chi^2_{min} = 0.31$.  For two degrees of freedom (four experiments
-- two parameters fitted), this is a very good fit comparable to the
case of the MSW solution in the mass mechanism where $\chi^2_{min} =
0.12$ (see \cite{SK}).  For the large mixing angle solution the fit is
considerably worse with $\chi^2_{min} = 3.4$.  The allowed regions
shown in Fig.~2 are compatible with the ones found in \cite{PHL}. The
differences can be attributed to the more recent experimental data
used in the present analysis as well as to the different ways in which
these data were treated.

We have repeated our analysis of the data by including the
gravitational field of the local supercluster which is estimated by
Kenyon\cite{pot} to be $3 \times 10^{-5}$.  This potential, being
three times the gravitational potential of the Sun at its center,
would dominate.  The allowed regions in this case change little in
shape but are shifted to lower values of $\Delta f$ by approximately a
factor of three.  The small mixing allowed region is shifted also to
smaller angles ($1.5 \times 10^{-3} \le \sin^2 (2\theta_G) \le 
6.0 \times 10^{-3}$) as the stronger gravitational field improves
adiabaticity if all the other parameters in Eq.~(\ref{eq:adia}) remain
the same.  The improved adiabaticity results in a broader suppression
pit in the survival probability and the pp-neutrinos become more
strongly suppressed for the same values of $\sin^22\theta_G$, which
comes into conflict with the results from the Gallium experiments.

Different components of the solar neutrino flux are suppressed
differently in the two allowed regions.  In the large mixing region
the pp-, $^7{\rm Be}$-, $^8{\rm B}$-, pep-, and CNO-neutrinos are
suppressed almost equally because the allowed $\Delta f$ values
correspond to the flat bottom part of the survival probability curve
(see Fig.~1b).  On the other hand, in the region of small mixing, the
$^7{\rm Be}$-neutrinos are suppressed more strongly than the rest of
the solar neutrinos as they are at the deepest part of the narrow
suppression pit.  This is similar to the analogous regions in the case
of a MSW solution for the mass mechanism.  However, there is an
important difference between the two cases, namely, the energy ranges
corresponding to adiabatic and nonadiabatic transitions are opposite.
It is this difference which leads to different neutrino spectra for
the two cases. The nonadiabatic edge in the case of the mass
mechanism is not as steep as the adiabatic edge in the case of the VEP
mechanism.  Since these are responsible for the spectral distortion of
the boron neutrinos one expects more abrupt changes of this spectrum
in the case of VEP.

\subsection{Spectral Distortion}

It has been shown in \cite{JB} that the solar neutrino spectrum is
independent of all solar model considerations to a very high
accuracy. Distortions of the spectra, if found experimentally, would
constitute a strong evidence for a neutrino physics solution of the
solar neutrino problem.

The Kamiokande-II (K-II) collaboration has obtained the first piece of
spectral information on solar neutrinos by measuring the energy
spectrum of the recoil electrons from neutrino-electron
scattering\cite{Kel}.  Because of the relatively large statistical
errors, the constraints on possible distortions of the spectrum are
not very stringent.  We have used the K-II data to rule out values of
the parameters $\Delta f$ and $\sin^2 2\theta_G$.  The excluded region
at 95 \% C.L. is shown as the hatched region in Fig.~2.  This
exclusion is obtained by comparing the predicted recoil--electron
spectrum with the measured one for a large number of values for
$\Delta f$ and $\sin^22\theta_G$, taking into account the energy
resolution and threshold efficiency function of the K-II detector.
The excluded region overlaps with part of the allowed ``small mixing
region" obtained from the $\chi^2$ analysis of the event rates
discussed above.  It should be emphasized that, while the position and
shape of the allowed regions in Fig.~2 depend on the predicted solar
neutrino flux from the standard solar model, the region excluded by
the recoil-electron spectrum is solar model independent.

The excluded region in Fig.~2 depends sensitively on the highest
energy data point in the K-II spectrum.  This point has a relatively
small error bar which is a result of combining the data from all
higher-energy bins above 13~MeV.  If this data point is ignored, we
find that the excluded region will be reduced considerably and will no
longer overlap with the allowed region.  The situation will be
improved when more precise measurements of the recoil-electron
spectrum become available from the upcoming SNO and Super-Kamiokande
detectors.

We have studied the possibility of using the SNO and Super-Kamiokande
measurements to identify the mixing mechanism responsible for the
solar neutrino problem.  We show in Fig.~3 the predicted spectra for
various allowed values of the mixing parameters.  Spectra predicted
for MSW transitions in the VEP mechanism and in the mass mechanism are
displayed in Fig.~3a and Fig.~3b, respectively.  What is actually
shown is the ratio of the predicted spectrum, ${\rm F(T_e)}$, to the
corresponding spectrum, ${\rm F_{st}(T_e)}$, calculated for the
standard solar model\cite{BP} with no neutrino mixing.  We normalize
the value of this ratio such that it is equal to 1 for
recoil-electrons with energy $T_e = 10$~MeV.  For large mixing angles,
there is little distortion from the standard solar model spectrum,
both for the VEP and for the mass mechanism.  This is why the measured
K-II spectrum only excludes a region corresponding to small mixing
angles (see Fig.~2).  On the other hand, the spectral distortion for
small mixing angles in the VEP case is noticeably different from that
in the mass mechanism.

One possible measure of the difference in the spectral distortion is
the derivative,
\begin{equation}
\xi_e ({\rm T_e}) = \frac{\rm d}{\rm dT_e} \left[\frac{\rm F(T_e)}
{\rm F_{st}(T_e)}\right]
\label{deriv}
\end{equation}
As an illustration of the sensitivity of this variable to the
distortions of the shape of the spectral curves we have compared its
values for the VEP and mass mechanisms at T$_{\rm e} = 10$~MeV.  
For the VEP mechanism $\xi_e (10~{\rm MeV})$ is equal to 0.31, 0.29
and 0.27 MeV$^{-1}$ for curves labeled from 1 to 3 in Fig.~3a.  In the
case of the mass mechanism the corresponding values are 0.036,
0.035 and 0.044 MeV$^{-1}$ for curves labeled 4 to 6 in Fig.~3b.  For
the rest of the curves shown in these two figures the derivative is
very close to zero, typically an order of magnitude smaller than the
above values.

Since long-wavelength vacuum oscillation for the mass mechanism is
still a possible solution to the solar neutrino problem, it is
necessary to also study the calculated recoil-electron spectra for
this case, which are shown in Fig.~3c.  We see that the spectral
distortion here can be as large as that in the VEP mechanism. However,
the character of this distortion is different. For curves labeled from
1 to 6 in Fig.~3c the corresponding values of $\xi_e (10~{\rm MeV})$ 
are -- 0.156, 0.0094, 0.057, 0.10, 0.089 and 0.0695 MeV$^{-1}$.
Furthermore, at energies below 10~MeV the general behavior of the 
spectral distortion is drastically different in the two cases.  Therefore 
future solar neutrino experiments should be able to distinguish, independent 
of solar model predictions of the neutrino fluxes, between the different
scenarios by comparing the shapes of the recoil--electron spectra,
provided the experimental uncertainties are sufficiently small.

In addition to neutrino-electron scattering, the SNO detector can also
detect the $^8{\rm B}$-neutrinos from the Sun by the process $\nu_e +
d \rightarrow p + p + e^-$.  In the case of the MSW effect for the
mass mechanism the $^8{\rm B}$-neutrino spectrum is smoothly and
almost uniformly distorted in the region between 5 and 14.5~MeV, which
is the interval of energies to which the SNO detector is expected to
be sensitive. On the other hand, in the case of the VEP mechanism, the
distortion of the $^8{\rm B}$-neutrino spectrum is much more abrupt in
the small mixing region.  This is illustrated in Figs.~4a and 4b where
the $^8{\rm B}$-neutrino spectra are shown in the two cases for sets
of allowed parameters.  Similar to Fig.~3, the spectra are normalized to 
the standard $^8{\rm B}$-neutrino spectrum (corresponding to no neutrino 
mixing) and to their values at 10~MeV.  It is evident from these figures 
that, for small mixing angles, the deviations from the standard $^8{\rm
B}$-neutrino spectrum are large.  Moreover, the spectral distortions
in the case of the VEP mechanism are strikingly different from the
corresponding ones in the case of the MSW solution for the mass
mechanism.  The derivative, $\frac{\rm d}{\rm dE_\nu} \left[{\rm
F(E_\nu)}/{\rm F_{st}(E_\nu)}\right]$, has values 1.1, 1.0 and 0.90 at $E_\nu
= 10$~MeV for curves 1 to 3 in Fig.~4a, whereas it is equal to 0.10,
0.093 and 0.10 for curves 4 to 6 in Fig.~4b.  The difference in the
shape of these curves is big enough to be measured in the SNO detector
as indicated by the estimated error bars after five years of operation
of this detector.\footnote{Efficiency and energy resolution have not
been included in the estimate of the error bars. The error bars are
simply the square root of the estimated number of events per year.}

We also compare the VEP spectra with those predicted for the mass
mechanism in the case of long-wavelength vacuum oscillations,
displayed in Fig.~4c.  The values of the corresponding derivative at
$E_\nu = 10$~MeV are -- 0.43, -- 0.16, 0.070, 0.23, 0.21 and 0.19
MeV$^{-1}$ for curves labeled 1 to 6 in Fig.~4c.  We see again a
measurable difference between the spectra in these two cases. The
neutrino oscillations in vacuum result in stronger distortions in the
lower spectrum of the energy interval between 5 and 14.5~MeV whereas
the VEP distortions are more prominent at the higher energies.

\section{Conclusion}

Using the current solar neutrino data, we find that the VEP mechanism
remains a viable solution to the solar neutrino problem.  The existing
data rule out a possible violation of the principle of equivalence for
a substantial region of the $\Delta f$ - $\sin^22\theta_G$ plane between
$10^{-18} < \Delta f < 10^{-12}$.  This result is much stronger than
the constraints obtained from SN1987A by comparing the arrival times
of neutrinos and photons, $|f_\nu - f_\gamma| < 3\times10^{-3}$
\cite{Longo}\cite{Krauss}, and by comparing neutrinos with
anti-neutrinos, $|f_\nu - f_{\bar{\nu}}| < 10^{-6}$
\cite{LoSecco}\cite{PSW}.  It is also stronger than the best limit 
of $10^{-12}$ derived from torsion balance experiments\cite{torsion},
which refers to macroscopic samples of matter and not to neutrinos.
Consequently, the violation of the equivalence principle by neutrinos,
indicated by the allowed regions in Fig.~2, does not translate into a
violation of the equivalence principle by the charged leptons at an
unacceptable level.  For example, it does not induce lepton flavor
changing transitions such as $\mu \rightarrow e \gamma$ at a rate
already excluded by experiment\cite{HLP}.

If the violation of the equivalence principle has the gradient form
given in Eq.~(\ref{eq:grad}) rather than the linear form in 
Eq.~(\ref{sub}), then from Eq.~(\ref{eq:scale}) and the limits
cited above for $\Delta f$, we conclude that a significant fraction of
flavor-violating couplings in the range $10^{-3}~{\rm eV}~
\moko~{\rm M}_f~\moko~10^3~{\rm eV}$ are excluded.

Observation of the distortion of the solar neutrino spectrum would be
a strong indication that neutrino physics is at the heart of the solar
neutrino problem.  We have shown that the recoil-electron spectrum
measured by Kamiokande-II excludes, in a solar model independent way,
a region of the otherwise allowed VEP parameter space.  We have
studied the prospects of using spectral measurements of solar
neutrinos to distinguish among various neutrino oscillation
mechanisms.  In the case of small mixing angles, spectral measurements
from upcoming solar neutrino experiments will be able to determine
which is the underlying mechanism of neutrino mixing.  On the other
hand, atmospheric neutrino data favor large mixing, in which case
spectral measurements of solar neutrinos cannot easily distinguish the
VEP mechanism from the mass mechanism.  In this case, long-baseline
accelerator neutrino experiments\cite{PHL}\cite{Iida} with typical
neutrino energies between 1 and 20 GeV and separations of order
hundreds of kilometers may provide the means to distinguish these two
mechanisms.

\vspace*{2.0 cm}
\noindent {\bf Acknowledgements}\\

We are grateful to A. M. Polyakov for an informative and stimulating
discussion of possible mechanisms for the violation of the equivalence
principle in string theories.  The work of P.I.K. has been partially
supported by Dyson Visiting Professor Funds from the Institute for
Advanced Study.  He would like to thank the Theory Group at Fermilab
and C.N.L. thanks the International School for Advanced Studies,
especially S. T. Petcov, for their hospitality during the latter stage
of this work.  C.N.L. also wishes to thank S. T. Petcov and A. Yu
Smirnov for useful discussions.  This work is supported in part by the
U.S. Department of Energy under Grants No.~DE--FG05--85ER--40219 and
No.~DE-FG02-84ER40163 and by the North Carolina Supercomputing
Program.

\raggedbottom

\newpage
\centerline{\bf Figure Captions}
\vspace*{1.0 cm}

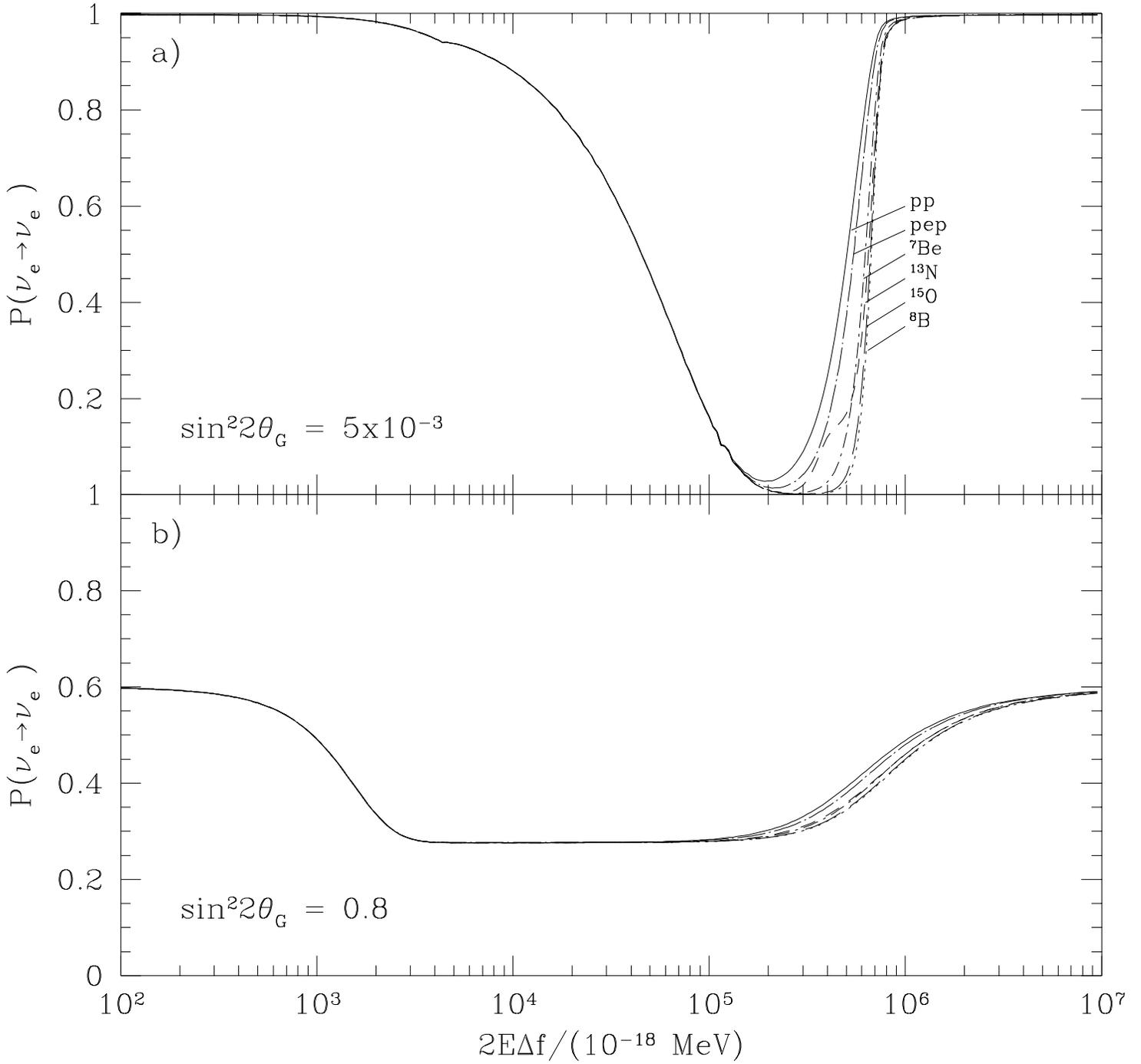
\noindent {\bf Fig.1}  Survival probabilties as a function of 
$E \Delta f $ for: a) $\sin^2 2 \theta_G = 5\times10^{-3}$ and b)
$\sin^2 2 \theta_G = 0.8$.  The different curves correspond to
averaging over the different neutrino production regions according to
the solar model in \cite{BP}.\\

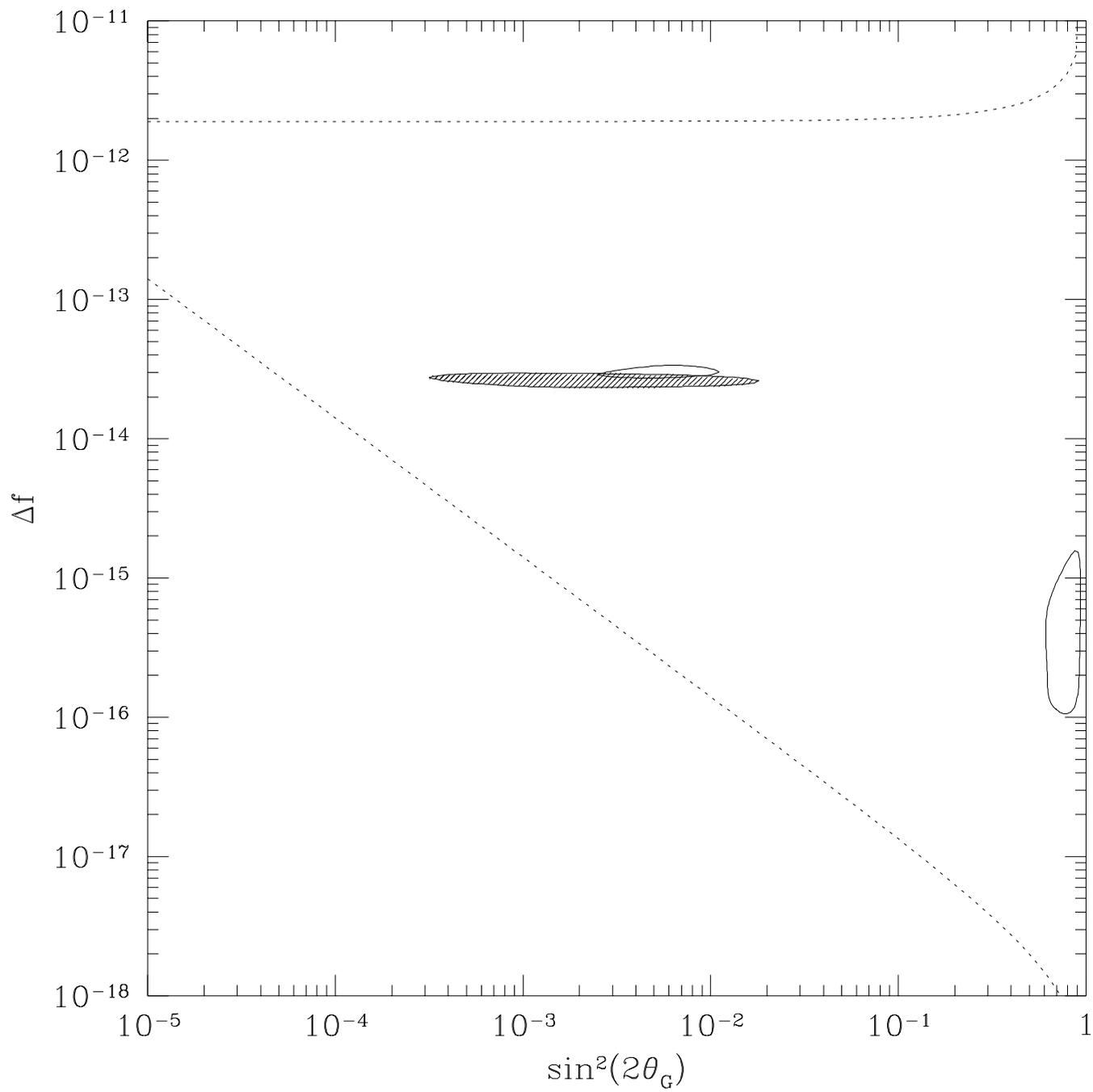
\noindent {\bf Fig.2}  95 \% C.L. allowed regions of the parameters 
$\Delta f$ and $\sin^2 2\theta_G$ derived from the latest solar
neutrino data (unhatched).  The region that can be probed with solar
neutrino experiments is bounded by the dotted lines.  The hatched
region is ruled out by the recoil--electron energy spectrum measured
by the Kamiokande--II collaboration.\\

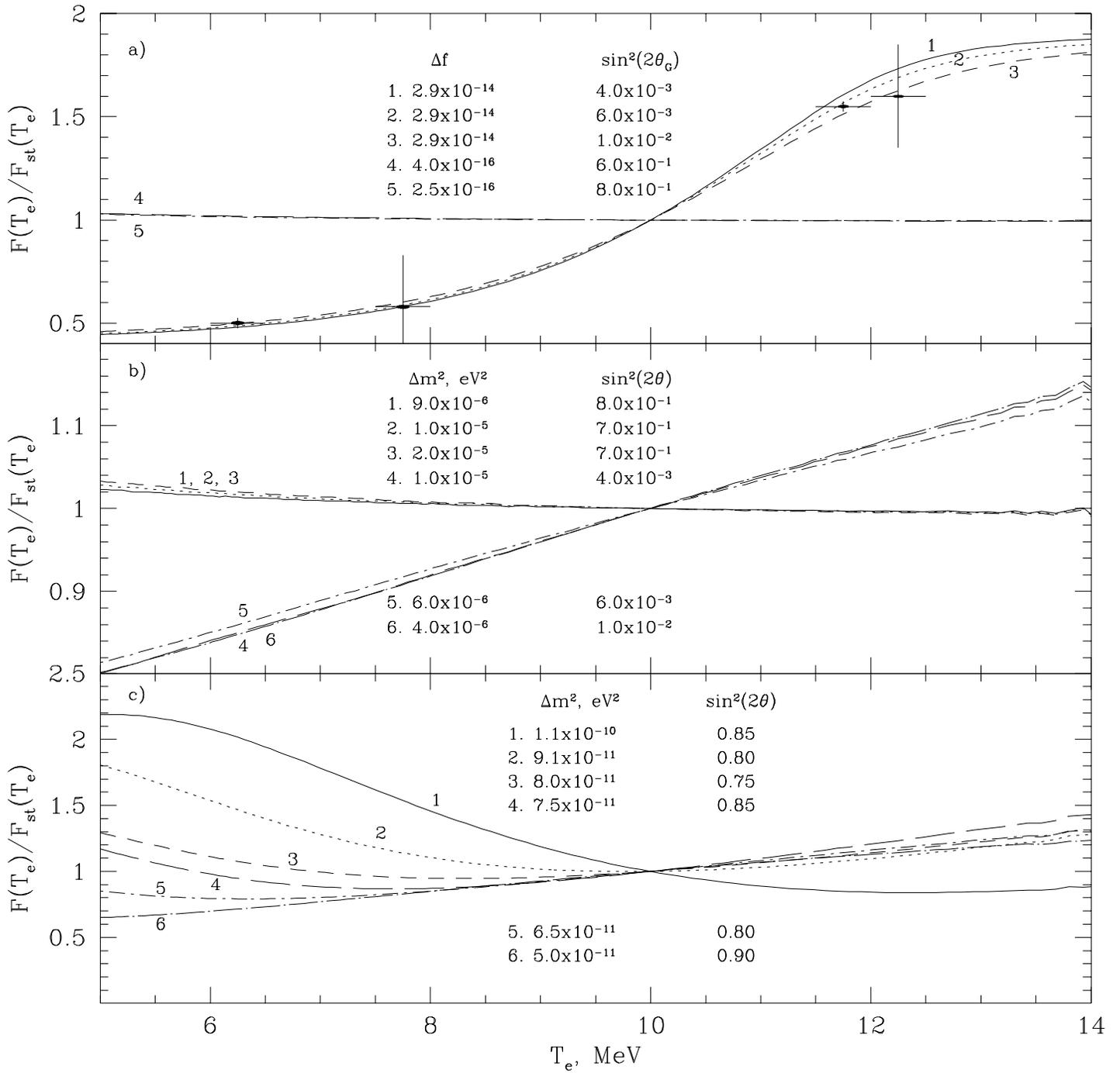
\noindent {\bf Fig.3}  Ratios of the predicted recoil-electron 
spectra for: a) ``MSW--enhanced" VEP mechanism, b) MSW effect in the
mass mechanism, and c) neutrino oscillations in vacuum (note the
different scale in each of these cases) to the standard spectrum of
recoil-electrons from boron neutirnos.  $T_e$ is the recoil-electron
energy.  The chosen values of the parameters for each case correspond
to values allowed by the current solar neutrino data.\\

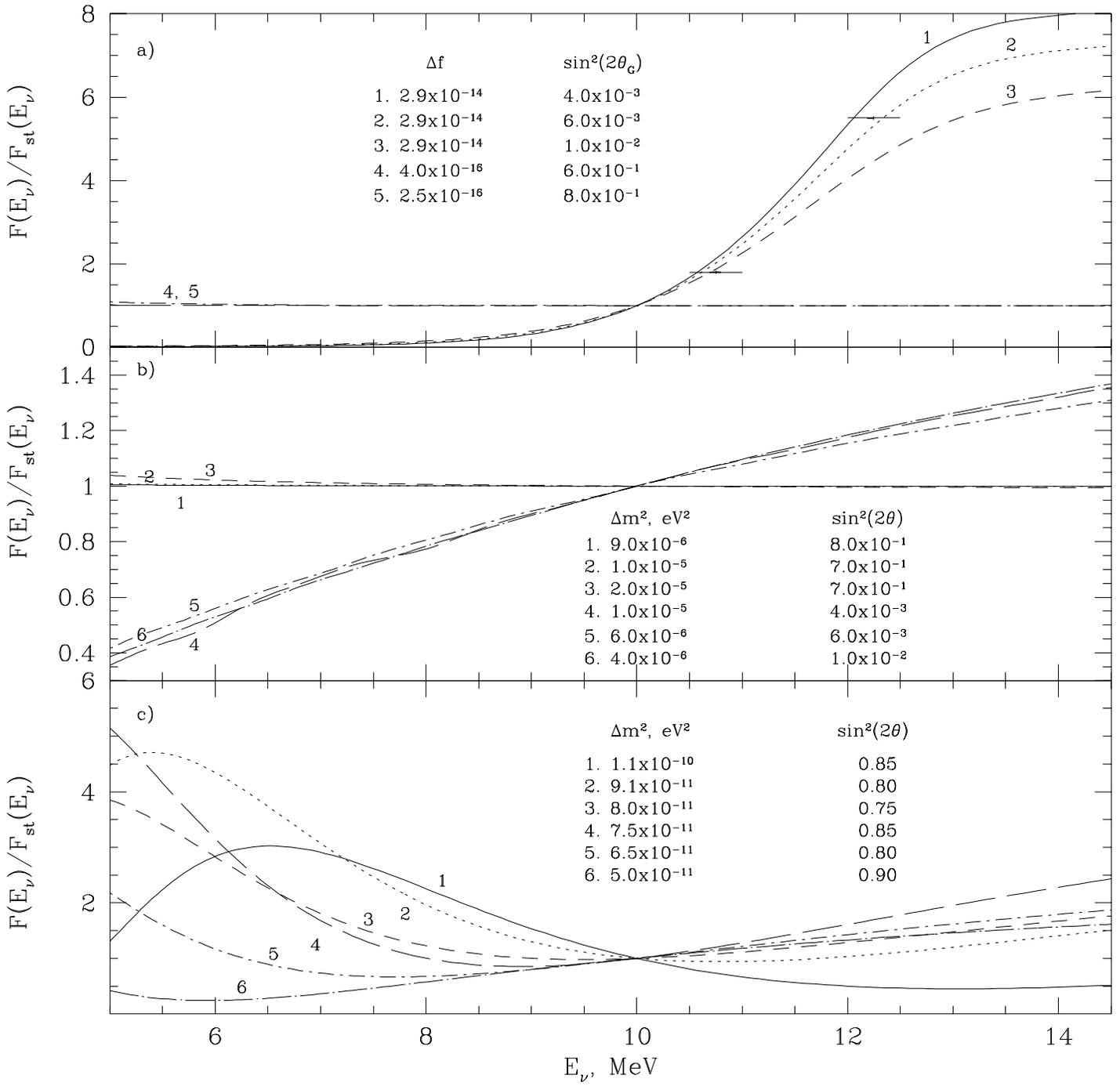
\noindent {\bf Fig.4}  Predicted ratios of $^8{\rm B}$-neutrino 
spectra to the standard boron neutrino spectrum for the same three
cases as in Fig.~3.  Note the different scale used in each case.
$E_\nu$ is the neutrino energy.

\end{document}